%% file: nD_thermalization.tex
\documentclass[%
nofootinbib,
superscriptaddress,
 preprint,
showpacs,preprintnumbers,
 amsmath,amssymb,
 aps,
pra,
 longbibliography,
 lengthcheck,%
]{revtex4-1}

\usepackage{nD_thermalization}

\usepackage{xcolor}
\hypersetup{
  colorlinks=true,
  linkcolor=blue!50!red,
  urlcolor=red!70!black
}

\begin{document}

\input{nD_thermalization_body.tex}

\end{document}

%% file: nD_thermalization_body.tex
\title{Friction as a consistent quantum-mechanical concept}

\author{Dmitry V. Zhdanov}
\email{dm.zhdanov@gmail.com}
\affiliation{University of Bristol, Bristol BS8\,1QU, UK}
\author{Denys I. Bondar}
\email{dbondar@tulane.edu}
\affiliation{Tulane University, New Orleans, Louisiana 70118, USA}
\author{Tamar Seideman}
\email{t-seideman@northwestern.edu}
\affiliation{Northwestern University, Evanston, Illinois 60208, USA}
\begin{abstract}
A quantum analog of friction (understood as a completely positive, Markovian, translation-invariant and phenomenological model of dissipation) is known to be in odds with the detailed balance in the thermodynamic limit. We show that this is not the case for quantum systems with internal (e.g. spin) states non-adiabatically coupled to translational dynamics. For such systems, a quantum master equation is derived which phenomenologically accounts for the frictional effect of a uniform zero temperature environment. A simple analytical example is provided. Conjectures regarding the finite temperature case are also formulated. The results are important for efficient simulations of complex molecular dynamics and quantum reservoir engineering applications.
\end{abstract}
\maketitle
\section{Introduction}

When dealing with complex dissipative environments, modern quantum scientists substantially rely on intuition, alike ancient craftsmen taming the elements to build mills or sailing vessels. Specifically, there is a vast experimental evidence that quantum effects play a pivotal role even in such complex and manifestly dissipative processes as photosynthesis \cite{2007-Engel,2014-Halpin,2018-Jang}. However, it not clear how coherent quantum dynamics is induced and guided by dissipative interactions. Despite several recent conceptual breakthroughs in the areas of quantum reservoir engineering%
\footnote{For instance, it was proven that quantum information processing can be fully dissipation-driven \cite{2013-Kastoryano}. Furthermore, dissipation enables optical and mechanical nonreciprocal couplings -- a key ingredient for implementing a quantum analog of sailing vessel \cite{2015-Metelmann}.} and topologically protected phases of matter \cite{BOOK-Tkachov-2015}, the analysis of coherent dynamics in real-would open systems is impeded by a prohibitively complex microscopic modelling of nonperturbative system-bath interactions. 

In classical mechanics, the similar curse of dimensionality is escaped from by introducing friction, a phenomenological non-conservative force resisting the relative motion of objects and converting their kinetic energies into heat \cite{BOOK-Razavy}. Despite being very simple, the concept of friction is proven useful even in explaining some quantum-level dynamics involving strong system-bath couplings, such as in the case of simple chemical reactions in liquid solutions \cite{2014-Orr-Ewing}. However, there is no unique way to quantize non-conservative forces \cite{BOOK-Razavy}. As result, existing phenomenological quantum dissipative models, be it quantum optical master equation \cite{BOOK-Gardiner}, F\"oster, or Redfield models \cite{1957-Redfield,2013-Jang}, depend on the system Hamiltonian in a complex way. This dependence cannot be simplified to few friction and diffusion coefficients, especially, in the low temperature regime. Furthermore, in order to ensure that the surrounding physical bath stays unchanged, quantum dissipation model must be nontrivially readjusted each time the Hamiltonian is altered (e.g., by an external field or as a result of chemical reactions). Furthermore, the discussed models describe the relaxation dynamics in terms of interstate transition rates. This picture is not natural when dealing with dynamics of essentially semiclassical vibration wavepackets in many photochemical reactions.

To mitigate the above complications, in this paper, we develop a phenomenological quantum dissipation model possessing favorable features of classical friction. For brevity, we will refer to this model as ``quantum friction''%
\footnote{We want to disambiguate our broad notion of ``quantum friction'' from its narrow meaning of the force acting on atoms flying near surfaces \cite{1999-Kardar}.}%
. We restrict our analysis to the simplest case of a homogeneous environment at a zero temperature and master equations linear with respect to the system density matrix. It is explained in detail elsewhere that the corresponding quantum friction model must satisfy the following four criteria \cite{2016-Bondar,2017-Zhdanov}:
\begin{enumerate}[i]
\item\label{__crit_M} Markovianity,
\item\label{__crit_P} positivity,
\item\label{__crit_TI} translation invariance, and
\item\label{__crit_E} asymptotic approach to the canonical equilibrium state.
\end{enumerate}
Specifically, the Markovianity ensures that the quantum friction is memoryless. Positivity guarantees that the model is quantum-mechanically consistent, i.e., that any initial positive density matrix remain positive at all times. Translation invariance makes quantum friction coordinate-independent, as the corresponding classical friction is a velocity-only dependent force.

The problem of finding a phenomenological model obeying these four criteria has a long history and has been the subject of many controversies over the years (see, e.g., Refs.~\cite{1998-Wiseman,2001-OConnell,2001-Vacchini}). As early as in 1976, Lindblad demonstrated that this problem has no resolution for a harmonic oscillator \cite{1976-Lindblad-a}. Subsequent failed searches forced Kohen et.al. to conjecture in their 1997 review \cite{1997-Kohen} that the incompatibility of all four criteria is a generic property, though with the sagacious comment: ``except in special cases'', which were not known at that time. 
Later, free quantum Brownian motion was identified by Vacchini \cite{2002-Vacchini} to be such a special case. We have proven that no other exceptions exist among quantum systems with $\Next$ translation degrees of freedom \cite{2017-Zhdanov}.

In this paper, we report a new class of ``special cases'' with all four criteria satisfied at zero temperature. These are the systems with \emph{internal} degrees of freedom, such as spin. The explicit forms of quantum friction dissipators are derived for such systems.

The paper is organized as follows. We start with formalizing the four-criterion problem in Sec.~\ref{@SEC:PS}. A constructive proof of existence of its solution at zero temperature for systems with internal degrees of freedom is presented in Section~\ref{@SEC:T=0}. The corresponding quantum friction dissipators are also derived and analyzed in detail. The obtained results are illustrated on the simplest analytically tractable example of a two-dimensional harmonic oscillator in Sec.~\ref{@SEC:Ill}. In the following section~\ref{@SEC:Dis} we discuss the conjectures regarding the existence of quantum friction at finite temperatures. The paper concludes with a brief summary and outlook.

\section{Formulation of the problem\label{@SEC:PS}}
The object of our analysis will be the quantum systems with $\Nint{>}1$ internal states $\ket{\es{i}}$ coupled to $\Next{\ne}0$ translation degrees of freedom. (Hereafter, the tilde $\tilde{~}$ marks the quantities associated with the internal states.) An example of such a system is a molecule with the following vibronic model Hamiltonian:
\begin{gather}
\hat H{=}\underbrace{\sum_{k{=}1}^{\Next}\frac1{2\mu_k}\hat p_k^2{+}V(\hat\xx)}_{\rm vibrational~part}{+}\underbrace{\sum_{i{=}1}^{\Nint}\ese{i}\proj{\es{i}}}_{\rm spin\backslash electronic~part}{+}\notag\\\underbrace{\sum_{i{\ne}j}
g_{i,j}\proj[\es{i}]{\es{j}}}_{\rm spin{\backslash}electronic~couplings}{+}\underbrace{\sum_{i{=}1}^{\Nint}\delta V_i(\hat\xx)\proj{\es{i}}}_{\rm vibronic~couplings}.\label{01.-H}
\end{gather} 
Here the sets of nuclear coordinates $\hat\xx{=}\{\hat x_1,...,\hat x_{\Next}\}$ and momenta $\hat\pp{=}\{\hat p_1,...,\hat p_{\Next}\}$ represent the translation degrees of freedom and ``internal'' states $\ket{\es{i}}$ correspond to different electronic and/or spin states of the molecule.
The symbols $g_{i.j}$, $V_i(\hat\xx)$ and $\ese{i}$ denote the coupling constants, coupling operators and internal states eigenenergies, respectively. Assume that the molecule is immersed into a homogeneous environment (such as gaseous media or uniform solvent) at zero temperature. Our goal is to construct a phenomenological model of the dissipative back-action of such environment on the system which satisfies the criteria \ref{__crit_M} -- \ref{__crit_E} of quantum friction.

The question which dissipative processes simultaneously satisfy criteria \ref{__crit_M} and \ref{__crit_P} was resolved by Lindblad \cite{1976-Lindblad}. The answer is a master equation now bearing his name:
\begin{gather}\label{01.-master_equation}
\tpder{}{t}\hat\rho{=}{\cal L}[\hat\rho],~{\cal L}{=}\Lvn{+}\Lrel,
\end{gather}
where the superoperator $\Lvn[\odot]{=}\tfrac{i}{\hbar}[\odot,\hat H]$ accounts for unitary evolution of the system isolated from environment and the dissipation term $\Lrel$ describing the system-environment couplings is
\begin{subequations}\label{01.-L_Markovian_translation_invariance}
\begin{gather}\label{01.-Lrel_definition}
\Lrel{=}\sum_{k{=}1}^{\Nlbd}\Lbd_{\hat A_k}
,\\
\label{01.-_Lindbladian_definition}
\Lbd_{\hat L}[\hat\rho]{\defeq}\hat L{\hat\rho}\hat L^{\dagger}{-}\tfrac12(\hat L^{\dagger}\hat L{\hat\rho}{+}{\hat\rho}\hat L^{\dagger}\hat L).
\end{gather}
The goal of this work is to identify the conditions under which the dissipator $\Lrel$ defined by Eq.~\eqref{01.-Lrel_definition} additionally obeys the criteria \ref{__crit_TI} and \ref{__crit_E}. 

The translation invariance criterion \ref{__crit_TI} with respect to the $n$-th coordinate is formally defined as
\begin{gather}\label{01.-L_translation_invariance}
[\hat p_n, \Lrel[\hat\rho]]{=}\Lrel[[\hat p_n, \hat\rho]].
\end{gather}
A general recipe on imposing this criterion on the Lindblad term \eqref{01.-Lrel_definition} was found by Holevo \cite{1995-Holevo,1996-Holevo} and further analyzed in applications by Vacchini \cite{2000-Vacchini,2005-Vacchini}. Specifically, to satisfy Eq.~\eqref{01.-L_translation_invariance} it is necessary and sufficient that the Lindblad operators $\hat A_k$ 
in Eq.~\eqref{01.-Lrel_definition} take the forms
\begin{gather}
\hat A_k{\defeq}e^{{-}i\kkappa_k\hat\xx}\hat\fc_k(\hat\pp)
.\label{01.-A_k}
\end{gather}
Here $\hat\fc_k(\pp){\in}\complexes^{\Nint{\times}\Nint}$ and $\kkappa_k{\in}\reals^{\Next}$ are arbitrary parameters and matrix-valued functions. They generally should be treated as a phenomenological quantities and can be deduced from empirical fits to time evolution of higher-order averages, e.g., $\midop{\hat{\xx}^2}$, $\midop{\hat{\pp}^2}$. In the cases when the dissipative term $\Lrel$ describes random collisions with light environmental particles, the values of $\hbar\kkappa_k$ and $\fc_k(\pp)$ can be associated with the characteristic momentum exchange in a collision event and the scattering amplitude in the momentum space \cite{2017-Zhdanov}.
\end{subequations}

It is worth to show why the dissipative dynamics satisfying criteria \ref{__crit_M} -- \ref{__crit_TI} deserves the name quantum friction. The average positions and momenta of a wavepacket, evolving according to Eqs.~\eqref{01.-L_Markovian_translation_invariance}, satisfy equations 
\begin{subequations}\label{ODM-eqs}
\begin{gather}
\label{ODM-dp/dt}
\tder{}{t}\midop{\hat p_n}{=}\frac{i}{\hbar}\midop{[\hat H,\hat p_n]} {+}\midop{\ehat\frnforce_n
},\\
\label{ODM-dx/dt}
\tder{}{t}\midop{\hat x_n}{=}\frac{i}{\hbar}\midop{[\hat H,\hat x_n]}{+}\midop{\ehat\xfrnforce_n},
\end{gather}
\end{subequations}
where
\begin{gather}\label{01.-friction_force}
{\ffrnforce}(\hat\pp){=}{-}\textstyle\sum_k\hbar\kkappa_{k}\fc_k^{\dagger}(\hat\pp)\fc_k(\hat\pp)
\end{gather}
defines the position-independent force $\midop{\ehat\ffrnforce}$ which, under a proper choice of operators $\fc_k^{\dagger}(\hat\pp)$, acts in the opposite direction to momenta $\midop{\hat{\pp}}$ similarly to the conventionally defined classical friction. Note, however, that this force is paired with term $\midop{\ehat\xxfrnforce}$ in Eq.~\eqref{ODM-dx/dt} where
\begin{gather}\label{01.-position_force}
{\xfrnforce_n}(\hat\pp){=}\frac12i\hbar\sum_k\left(
\fc_{k}^{\dagger}(\hat\pp)\tpder{\fc_k(\hat\pp)}{\hat p_n}{-}\tpder{\fc_k^{\dagger}(\hat\pp)}{\hat p_n}\fc_k(\hat\pp)
\right).
\end{gather}
The presence of term $\midop{\ehat\xxfrnforce}$ can be physically attributed to changed effective masses of moving particles ``dressed'' by the environment. This term is of the same nature as the ``position diffusion'', a well-known peculiarity of quantum Brownian motion \cite{2009-Vacchini}.

Recall that the classical frictional forces are accompanied by thermal fluctuations satisfying the fluctuation-dissipation theorem. Similarly, criteria \ref{__crit_M} -- \ref{__crit_TI} are not sufficient to define thermodynamically consistent quantum friction, and the additional criterion \ref{__crit_E} is required to ensure detailed balance at the thermal equilibrium: 
\begin{gather}\label{01.-L_thermalization}
\Lrel[\rhoth{T}]{=}0.
\end{gather}
Here $\rhoth{T}{=}\left.\hat\rho\right|_{t\to\infty}{\propto}e^{-\frac{\hat H}{k_{\mathrm{B}}T}}$ is the stationary Gibbs system state corresponding to temperature $T$. It is worth stressing that Eq.~\eqref{01.-L_thermalization} implies non-vanishing thermal fluctuations even at the zero bath temperature.

The criteria \ref{__crit_M}, \ref{__crit_P} and \ref{__crit_E} are fulfilled, e.g., by quantum optical master equation \cite{BOOK-Gardiner}. A variety of other models are known where some three out of four criteria \ref{__crit_M} -- \ref{__crit_E} are satisfied (see Ref.~\cite{1997-Kohen} for detailed review). However except for free Brownian motion \cite{2002-Vacchini}, no model is known which satisfies all the Eqs.~\eqref{01.-L_Markovian_translation_invariance} and \eqref{01.-L_thermalization}. Non-existence of such model for the case $\Nint{=}1$ was rigorously proven by us recently \cite{2017-Zhdanov}. In particular, we have shown that the criteria \ref{__crit_M}, \ref{__crit_P} and \ref{__crit_TI} are compatible only with the following weaker variant of the condition \eqref{01.-L_thermalization}:
\begin{gather}\label{01.-L_therm_approx}
\Tr[\rhoth{T'}\Lrel[\rhoth{T}]]{=}0 \mbox{ for any } T',
\end{gather}
which we will refer to as the \emph{relaxed thermalization} (RT) condition.

The condition \eqref{01.-L_therm_approx} always holds when Eq.~\eqref{01.-L_thermalization} is satisfied. In fact, it can be shown that Eq.~\eqref{01.-L_therm_approx} only guarantees that the steady state of the model coincides with the true equilibrium $\rhoth{T}$ up to the first order in system-bath couplings. Thus, Eq.~\eqref{01.-L_therm_approx} is expected to reasonably well approximate the exact thermalization criterion \eqref{01.-L_thermalization} when the bath-induced decay and decoherence times are large compared to all the characteristic dynamical timescales of the system \cite{2017-Zhdanov}.

\section{Reconciling the four-criterion clash at zero temperature
\label{@SEC:T=0}} 
In this section, we prove the existence of quantum friction simultaneously satisfying criteria \ref{__crit_M} -- \ref{__crit_E} for systems characterized by the Hamiltonian \eqref{01.-H} with $\Nint{>}1$. Our methodology is to first figure out the necessary requirements to satisfy the criteria \ref{__crit_M} -- \ref{__crit_TI} and RT criterion \eqref{01.-L_therm_approx} for $\Nint{>}1$ and then to identify whether the RT criterion can be upgraded to the exact thermalization condition \eqref{01.-L_thermalization}.

As discussed in the previous section, the dissipator $\Lrel$ satisfying criteria \ref{__crit_M} -- \ref{__crit_TI} needs to have the form defined by Eqs.~\eqref{01.-L_Markovian_translation_invariance}. In the case $T{=}0$, substitution of Eqs.~\eqref{01.-L_Markovian_translation_invariance} allows to cast Eq.~\eqref{01.-L_therm_approx} into $\Nlbd$ independent extremal conditions
\begin{gather}\label{01.-L_therm-A_k(T=0)}
J_k{=}\Tr[\rhoth{0}\Lbd_{\hat A_k}[\rhoth{0}]]{\to}\max~~(k{=}1,...,\Nlbd).
\end{gather}
Specifically, Eq.~\eqref{01.-L_therm-A_k(T=0)} can be obtained from Eq.~\eqref{01.-L_therm_approx} and equalities $\Tr[\Lbd_{\hat A_k}[\hat\rho]]{=}0$ which guarantee that $J_k{\leq}0$ for any $k$. Hence, Eq.~\eqref{01.-L_therm_approx} can be satisfied only when all $J_k$ take their maximal values $J_k{=}0$.

Let $\ket{n}$ and $E_n$ be the system Hamiltonian eigenstates and eigenvalues, respectively. Each of these states can be represented in the form
\begin{gather}
\ket{n}{=}\sum_{i{=}1}^{\Nint}\ket{\es{i}}\ket{\Psi_{n,i}},
\end{gather}
where $\ket{\Psi_{n,i}}$ are the vibrational parts of the eigenstates. Note that $\ket{\Psi_{n,i}}$ are neither normalized nor orthogonal. Below we will deal with their momentum wavefunctions $\Psi_{n,i}(\pp)$ and also with the operators $\Psi_{n,i}(\hat\pp)$ obtained via the substitution $\pp{\to}\hat\pp$. In addition, we will use the notation $\ket{\varphi_{k,n}}{=}\hat A_k\ket{n}$. The expressions for $J_k$ in Eq.~\eqref{01.-L_therm-A_k(T=0)} can now be rewritten as
\begin{gather}\label{01.-L_therm-A_k(T=0)'}
J_k{=}|\scpr{0}{\varphi_{k,0}}|^2{-}\scpr{\varphi_{k,0}}{\varphi_{k,0}}.
\end{gather}
Here we accounted for the fact that the thermodynamic equilibrium at $T{=}0$ corresponds to the ground state $\rhoth{0}{=}\proj{0}$. The Cauchy-Schwarz inequality requires that
\begin{gather}
|\scpr{0}{\varphi_{k,0}}|^2{\leq}\scpr{\varphi_{k,0}}{\varphi_{k,0}}\scpr{0}{0}{=}\scpr{\varphi_{k,0}}{\varphi_{k,0}},
\end{gather}
where the equality holds iff $\ket 0{\propto}\ket{\varphi_{k,0}}$. Hence, it follows from Eqs.~\eqref{01.-L_therm-A_k(T=0)} and \eqref{01.-L_therm-A_k(T=0)'} that
\begin{gather}\label{01.-cond_max(T=0)}
J_k{=}0\mbox{ iff }\hat A_k\ket 0{=}\alpha_k\ket 0 \mbox{ for all }k~~(\alpha_k\in{\mathbb C}).
\end{gather}
Eqs.~\eqref{01.-cond_max(T=0)} always can be resolved with respect to $\hat A_k$. The general solution can be written in terms of operators $\hat\fc_k(\hat{\pp})$ introduced in Eq.~\eqref{01.-A_k} as
\begin{subequations}\label{01.-solution_i+ii+iii+RT}
\begin{gather}\label{01.-fc_k-sol-v.2}
\hat\fc_k(\hat{\pp}){=}\hat \fc_{0,k}(\hat{\pp}){+}\alpha_k\sum_{i{=}1}^{\Nint}\frac{\Psi_{0,i}(\hat\pp{-}\hbar\kkappa_k)}{\Psi_{0,i}(\hat\pp)}\proj{\es i},
\end{gather}
where $\hat \fc_{0,k}(\hat\pp)$ is the position-independent operator satisfying the equation
\begin{gather}
\hat \fc_{0,k}(\hat\pp)\ket 0{=}0.
\end{gather}
\end{subequations}
For instance, the general solution for $\hat \fc_{0,k}$ in the case of a two-level internal subsystem $(\Nint{=}2)$ can be represented as
\begin{align}
\ehat\fc_{0,k}(\hat{\pp}){=}&\sum_{i,j{=}0}^1(-1)^{i{-}j}G_{1{-}i,k}(\hat\pp)\Psi_{0,i}^*(\hat\pp)\Psi_{0,j}(\hat\pp)\proj[\es{1{-}i}]{\es{1{-}j}},
\label{01.-fc_{0,k}-sol-v.2}
\end{align}
where 
$G_{0,k}(\pp){\in}\complexes$ and $G_{1,k}(\pp){\in}\complexes$ are arbitrary functions. 

One can notice that the operators \eqref{01.-fc_{0,k}-sol-v.2} always contain off-diagonal terms between different internal states $\ket{\es{i}}$. Hence, the population of internal states is conserved when $\hat \fc_{0,k}{=}0$ for all $k$. In the case when $\ket{\es{i}}$ represent the electronic states of a molecule, the latter model can represent instantaneous events (e.g., the direct collisions of light particles with the molecule's nuclei) not involving electrons. 

Equations \eqref{01.-solution_i+ii+iii+RT} answer the question on when criteria \ref{__crit_M} -- \ref{__crit_TI} can be satisfied together with the relaxed thermalization condition \eqref{01.-L_therm_approx}. Let us now turn to the conditions required to satisfy strict thermalization condition \ref{__crit_E} (Eq.~\eqref{01.-L_thermalization}). The substitution of Eqs.~\eqref{01.-solution_i+ii+iii+RT} allows to rewrite condition \eqref{01.-L_thermalization} as
\begin{gather}\notag
\sum_{k{=}1}^{\Nlbd}{\alpha_k^2}\ket{0}{=}\sum_{k{=}1}^{\Nlbd}\fc_k(\hat\pp)^{\dagger}\fc_k(\hat\pp)\ket{0}{=}\sum_{k{=}1}^{\Nlbd}\alpha_kA_k^{\dagger}(\hat\pp)\ket{0}
.
\end{gather}
It is easier to analyze this relation after multiplying both its sides by operator $\sum_{j{=}0}^{\Nint}\proj[\es{0}]{\es{j}}\Psi_{0,j}(\hat\pp)$, which gives
\begin{gather}\label{01.-cond_therm_strong}
\forall\pp:\sum_{k{=}1}^{\Nlbd}{\alpha_k^2}\sum_{j{=}1}^{\Nint}\left(\left|\Psi_{0,j}(\pp)\right|^2{-}\left|\Psi_{0,j}(\pp{-}\hbar\kkappa_k)\right|^2\right){=}0.
\end{gather}
As discussed in Ref.~\cite{2017-Zhdanov}, the effect of the part of Lindblad operator $A_k(\hat\pp)$ proportional to $\alpha_k$ can be associated with an instantaneous inelastic collision with massless particle, such as photon, having momentum $\hbar\kkappa_k$. In light of this interpretation, the exact thermalization condition \eqref{01.-cond_therm_strong} requires invariance of the momentum distribution with respect to entire sequence of such collisions described by operator $\Lrel$. Apart from exceptional cases, the condition \eqref{01.-cond_therm_strong} can be satisfied iff either $\alpha_k{=}0$ or $\kkappa_k{=}0$ for each $k$. Note, however, that the terms $\Lbd_{\hat A_k}$ corresponding to $\kkappa_k{=}0$ do not contribute to friction $\ehat\ffrnforce$ in Eq.~\eqref{ODM-dp/dt}. In other words, despite the underlying dissipative process formally satisfies criteria \ref{__crit_M} -- \ref{__crit_TI} of quantum friction, it does not involve direct momentum transfer between the system and the bath. Hence, it cannot be physically interpreted as a frictional process and, thus, is out of scope for our programme. 

The remaining possibility to obey condition \eqref{01.-cond_therm_strong} by setting $\alpha_k{=}0$ and $\kkappa_k{\ne}0$ in Eqs.~\eqref{01.-solution_i+ii+iii+RT} leads to exactly thermalizable dissipator $\Lrel$ satisfying Eq.~\eqref{01.-L_thermalization}. This finding that all four criteria \ref{__crit_M} -- \ref{__crit_E} can be simultaneously satisfied is the central result of this work. For example, in the case of the two-level internal subspace considered above, this is achieved by setting $\alpha_k{=}0$ in Eq.~\eqref{01.-fc_k-sol-v.2} and choosing arbitrary functions $G_{0,k}(\pp)$, $G_{1,k}(\pp)$ and vectors $\kkappa_k$ in Eq.~\eqref{01.-fc_{0,k}-sol-v.2}. It is worth to stress, however, that unlike classical friction, this solution leads to non-vanishing term $\midop{\xxfrnforce}$ in Eq.~\eqref{ODM-dx/dt}.

What makes quantum friction possible in the case $\Nint{>}1$? Our formal results admit the following physical interpretation. The very notion of friction is implicitly attached to the classical concept of bath. When considering classical dynamics, it is sufficient to treat the bath as an infinite heat tank at a constant temperature. However, this model fails to account for proximity effects responsible for spatial and/or temporal system-bath correlations. These correlations turned out to be crucial for quantum thermalization \cite{2017-Zhdanov}: Without them, microscopic perpetual motion would be possible. Internal degrees of freedom enable quantum friction by serving as an ancilla subsystem to phenomenologically mimic proximity effects on translation degrees of freedom. We have seen that the thermalizability criterion \ref{__crit_E} can be fulfilled even if this ancilla subsystem consists of just two quantum states. However, a very essence of proximity effects implies that this mechanism can work only when the external and ancilla internal degrees of freedom are coupled non-adiabatically. This implies that the system ground state $\ket{0}$ is such that $\ket{\Psi_{0,i}}{\not\propto}\ket{\Psi_{0,j{\ne}i}}$ and $\ket{\Psi_{0,i}}{\ne}0$ for all $i$ and $j$. If this condition is satisfied, quantum friction simultaneously thermalizes \emph{both} the external degrees of freedom and the ancilla subsystem. Otherwise, neither of these degrees of freedom can be thermalized, as can be seen, e.g., from Eq.~\eqref{01.-fc_{0,k}-sol-v.2} in the case of $\ket{\Psi_{0,1}}{=}0$.

\section{An illustration: the damped harmonic oscillator \label{@SEC:Ill}}
Let us illustrate the conclusions of the previous section by quantizing the familiar classical model of a damped harmonic oscillator
\begin{gather}\label{02.-classical_damped_oscillator}
\ddot{x}_1{+}2\gamma(\dot x)\dot x{+}\omega_1^2x_1{=}0,
\end{gather}
where $\omega_1$ and $\gamma{=}\gamma(\dot x)$ are the oscillator's frequency and frictional damping rate, respectively. Recall that according to celebrated Lindblad's result \cite{1976-Lindblad-a}, the model \eqref{02.-classical_damped_oscillator} cannot be quantized within criteria \ref{__crit_M} -- \ref{__crit_E} imposed on the friction term under the assumption that an oscillating body is a structureless point particle. Here we assume that the oscillating body is a prealigned diatomic molecule AB of mass $M$ with one internal harmonic degree of freedom: a stretching vibrational mode $x_2$ characterized by the reduced effective mass $\mu$ and potential energy $\hat V_{\rm{vib}}{\propto}\hat x_2^2$.

\begin{figure}[tbp]
\centering\includegraphics[width=0.99\columnwidth]
{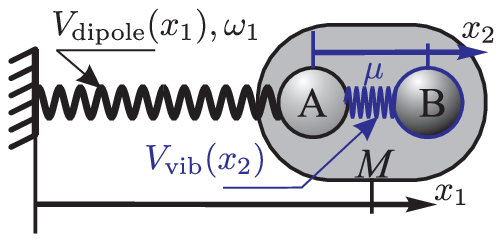}
\caption{The ball-and-spring model of a quantum damped harmonic oscillator (Eq.~\eqref{02.-classical_damped_oscillator}) with an additional internal structure represented by the molecular stretching mode $x_2$ (See Sec.~\ref{@SEC:Ill} for details).
\label{@FIG.01}}
\end{figure}

Consider the dynamics of such a molecule in a harmonic dipole trap, as shown in Fig.~\ref{@FIG.01}, assuming that the long-term thermalization dynamics is guided by an effective friction force (e.g., due to radiation decay), which stirs the system into the thermodynamic equilibrium with an environment. Assuming that atom A is a primary contributor to the induced dipole moment, the system Hamiltonian can be written in the form:
\begin{gather}\label{02.-H_mol_trap}
\hat H{=}\frac{\hat p_1^2}{2M}{+}\frac{\hat p_2^2}{2\mu}{+}\hat V_{\rm{dipole}}{+}\hat V_{\rm{vib}},\notag
\end{gather} 
where $x_1$ in an external center-of-mass molecular coordinate and $\hat V_{\rm{dipole}}{\simeq}\frac{M\omega_1^2}2(\hat x_1{-}\frac{\mu}{m_1}\hat x_2)^2$ is a laser-induced trapping potential. It is convenient to rewrite the Hamiltonian \eqref{02.-H_mol_trap} in the normal mode representation:
\begin{gather}\label{02.-H_vibr}
\hat H{=}\sum_{l=1}^2\hbar\omega_l{\hat a_l'}^{\dagger}\hat a_l'.
\end{gather}
Here $\hat a_l'$ is the annihilation operator for $l$-th normal mode:
\begin{gather}
\hat a_l'{=}\frac1{\sqrt2}\left(\frac1{\hbar\sqrt{\beta_l'}}\hat x_l'{+}i\sqrt{\beta_l'}\hat p_l'\right)~~(\beta_l'{=}(\hbar 
\omega_l)^{-1}),
\end{gather}
and the operators $\hat x_l'$ and $\hat p_l'$ of normal coordinates and momenta are defined as
\begin{align}
\hat x_1'&{=}\sqrt{m_1}\cos(\hophase)\hat x_1{-}\sqrt{m_2}\sin(\hophase)\hat x_2,\notag\\
\hat x_2'&{=}\sqrt{m_1}\sin(\hophase)\hat x_1{+}\sqrt{m_2}\cos(\hophase)\hat x_2,\notag\\
\hat p_1'&{=}\tfrac{\cos(\hophase)}{\sqrt{m_1}}\hat p_1{-}\tfrac{\sin(\hophase)}{\sqrt{m_2}}\hat p_2,~~
\hat p_2'{=}\tfrac{\sin(\hophase)}{\sqrt{m_1}}\hat p_1{+}\tfrac{\cos(\hophase)}{\sqrt{m_2}}\hat p_2.
\end{align}
We are interested in the situation when the external and internal motions are coupled, i.e., when $\hophase{\ne}0,\pi$. 

Let us derive the phenomenological frictional dissipator for the case of cold environment $T{=}0$. According to the conclusions of Sec.~\ref{@SEC:T=0}, the dissipator of interest should be of the form \eqref{01.-L_Markovian_translation_invariance} with
\begin{gather}
\hat A_k{=}e^{-i\kappa_k \hat x_1}f_k(\hat p_1,\hat p_2, \hat x_2)
\end{gather}
and satisfy Eq.~\eqref{01.-cond_max(T=0)}.

It is helpful to introduce the operators
\begin{gather}
\hat{\frak{a}}_1{=}{\frak{a}}_1(\hat x_1,\hat p_1,\hat p_2){=}\sqrt{2} \left(
\hat{a}'_2 \sqrt{\beta_2'} \sin (\hophase){+}\hat{a}'_1 \sqrt{\beta_1'} \cos(\hophase)
\right),\notag\\
\hat{\frak{a}}_2{=}{\frak{a}}_2(\hat x_2,\hat p_1,\hat p_2){=}\sqrt{2} \left(
\hat{a}'_2 \sqrt{\beta_2'} \cos (\hophase){-}\hat{a}'_1 \sqrt{\beta_1'} \sin(\hophase)
\right),
\end{gather}
which satisfy
\begin{gather}
\hat{\frak{a}}_{1,2}\ket{0}{=}0.
\end{gather}
Here $\ket 0$ is the system's ground state and also the equilibrium Gibbs state for $T=0$. It can be written in momentum representation as
\begin{gather}
\scpr{p_1,p_2}{0}{\propto}e^{-\sum_{l{=}1}^2\frac1{2}{\beta_l' p_l'^2}}.
\end{gather}

The general solution of the equation $\hat A_k\ket{0}{=}\alpha_k\ket{0}$ compliant with the translation invariance condition \eqref{01.-L_translation_invariance} with respect to $x_1$ is
\begin{gather}\label{03.-A_k-harm(T=0)}
\hat A_k{=}e^{-i\kappa_k\hat{x}_1}\hat{G}_k'\hat{\frak{a}}_2{+}\alpha_k e^{-i\hbar\kappa_k\frac{1}{\sqrt{m_1}}\hat{\frak{a}}_1}.
\end{gather}
Here $\hat G'_k{=}G'_k(\hat p_1,\hat p_2, \hat x_2)$ is an arbitrary operator independent of $\hat x_1$. Eq.~\eqref{03.-A_k-harm(T=0)} is nothing but the specialization of the general solution \eqref{01.-solution_i+ii+iii+RT}. It obeys the exact thermalization condition \eqref{01.-L_thermalization} if $\alpha_k{=}0$ for all $k$ and satisfies only RT condition \eqref{01.-L_therm_approx} otherwise. 

As expected, the dissipator $\Lrel$ corresponding to solution \eqref{03.-A_k-harm(T=0)} with $\alpha_k{=}0$ vanish when the internal and external degrees of freedom are dynamically decoupled (e.g., when $\Nint{=}1$ or $\hophase{=}0$). However, the solutions of both Eqs.~\eqref{01.-fc_k-sol-v.2} and \eqref{03.-A_k-harm(T=0)} with $\alpha_k{\ne}0$ do exist even in the latter case. A simple computation shows that these solutions reduce to $\fc_{k}(\hat p_1){\propto}e^{\frac{\hat p_1\kappa_k}{m_1\omega_1}}$ obtained in Ref.~\cite{2017-Zhdanov}. The latter solution (and, more generally, the last term in Eq.~\eqref{03.-A_k-harm(T=0)}) indicate that the system-environment correlations are minimized (for any given $\kkappa$) when the associated effective force $\midop{\xxfrnforce_k}$ depends exponentially on momenta $\hat{\pp}$. Interestingly that unlike typical classical friction forces anti-aligned with momenta, the last term in Eq.~\eqref{03.-A_k-harm(T=0)} has a long exponentially vanishing ``endothermic'' tail representing a force aligned with $p_1$. For a further discussion on the physics of these tails see Ref.~\cite{2017-Zhdanov}.

In contrast, the solution \eqref{03.-A_k-harm(T=0)} with $\alpha_k{=}0$ represents an arbitrary nonlinear friction force acting on the external degree of freedom $x_1$. Importantly, this force exists due to the dissipative coupling between the external and internal degrees of freedom. The origin for this coupling can be illustrated by the classical ball-and-spring model of a diatomic molecule where each ball is subjected to an independent non-linear friction force. In this model, the total effective friction force applied to the system's center of mass depends on the relative velocity of the balls, whereas the decay of the internal oscillation depends on the center-of-mass velocity. The hallmark of the quantum friction is inability to cancel these interdependencies out even for the linear friction $\frnforce_1{\propto}p_1$.

\section{Discussion on the finite temperature case\label{@SEC:Dis}}

So far, our analysis has been restricted to interactions of quantum systems with baths cooled down to the zero temperature. An existence of quantum friction forces at finite temperatures is an open question beyond the scope of this work. Nevertheless, we would like to briefly discuss insights that might help to find the answer.

First, note that in the case $T{=}0$ the exactly thermalizable frictional dissipator (defined by Eqs.~\eqref{01.-L_Markovian_translation_invariance} and \eqref{01.-solution_i+ii+iii+RT} with $\alpha_k{=}0$) has the property that each individual term $\Lbd_{\hat A_k}$ independently satisfies the condition \eqref{01.-L_thermalization}. However, the analogous property cannot hold at finite temperatures for any Lindblad operator $\hat A_k$ of the form \eqref{01.-A_k}:
\begin{gather}\label{04.-no-go_App_1}
\forall T{\ne}0: \Lbd_{\hat A_k}[\rhoth{T}]{\ne}0.
\end{gather}
The proof of inequality \eqref{04.-no-go_App_1} is given in Appendix~\ref{@APP:1}. This result can be intuitively understood in the simplest case $\Next{=}1$ as follows. The most general form of the operator $\hat\fc_k$ in Eq.~\eqref{01.-A_k} corresponding to a single translational degree of freedom is
\begin{gather}
\hat\fc_k{=}\sum_{i,j=1}^{\Next}\sum_{k}\proj[\es{i}]{\es{j}}c_{i,j,k}\hat p^k,
\end{gather}
where $c_{i,j,k}$ are complex coefficients to be determined. Now, imagine that we truncated the translational basis to $K$ states. It is obviously impossible to satisfy the exact thermalization condition in this approximation. Indeed, in order to turn the inequality \eqref{04.-no-go_App_1} into the equality while satisfying the condition \eqref{01.-L_thermalization}, $(\Nint K)^2-1$ constraints must be satisfied with only $\Nint^2K{-}1$ unknowns $c_{i,j,k}$ (here we excluded the complex scaling factor).

As discussed in Ref.~\cite{2017-Zhdanov}, the terms $\Lbd_{\hat A_k}$ can be regarded as Markovian approximations for relaxation processes similar to ones involved in the Doppler cooling. Importantly, each term $\Lbd_{\hat A_k}$ represents an independent frictional process in such an interpretation. At the same time, the inequality \eqref{04.-no-go_App_1} shows that these processes cannot be independent since each of them drives the system out of thermal equilibrium. This contradiction shows that if the friction-like Markovian dissipator $\Lrel$ exists for finite temperatures, it must consist of several interdependent terms of the form $\Lbd_{\hat A_k}$, and hence has a non-trivial physical interpretation, as in non-TI case \cite{1976-Alicki}. Alternatively, the contradiction may signify that no friction-like quantum process exists at nonzero temperatures. Nevertheless, if the latter conjecture is correct, the conventional interpretation of friction in the classical limit would require revisiting.
 
At the same time, the RT condition
\begin{gather}\label{04.-go_RT}
\forall T': \Tr[\rhoth{T'}\Lbd_{\hat A_k}[\rhoth{T}]]{=}0
\end{gather}
can be seamlessly satisfied. Indeed, Eq.~\eqref{04.-go_RT} sets only $\Nint K{-}1$ constraints (which is less than the previously mentioned number of $\Nint^2K{-}1$ parameters) and can be obeyed together with the TI condition \eqref{01.-L_translation_invariance}. For instance, the dissipation superoperator $\Lrel$ for the harmonic oscillator example of Sec.~\ref{@SEC:Ill} satisfying criteria \ref{__crit_M} -- \ref{__crit_TI} and RT condition \eqref{04.-go_RT} at a finite temperature is given by Eqs.~\eqref{01.-L_Markovian_translation_invariance} with
\begin{gather}
\hat A_k{\propto}e^{-i \kappa_k \frac{\left(\hat{x}'_2 \sin (\hophase)+\hat{x}'_1 \cos (\hophase)\right)}{\sqrt{m_1}}}
e^{\kappa _k \frac{\left(\beta _2' \lambda _2 \hat{p}'_2 \sin (\hophase)+\beta _1' \lambda _1 \hat{p}'_1 \cos (\hophase)\right)}{\sqrt{m_1}}}{=}\notag\\
e^{-i \kappa_k \hat x_1}
e^{\kappa _k \frac{1}{\sqrt{m_1}}\left(\beta _2' \lambda _2 \hat{p}'_2 \sin (\hophase)+\beta _1' \lambda _1 \hat{p}'_1 \cos (\hophase)\right)},
\end{gather}
where $\lambda_l{=}\tanh(\frac{\hbar\omega_l}{4k_{\rm B}T})$.

\section{Conclusion}

The concept of friction (defined as the phenomenological dissipative model satisfying criteria \ref{__crit_M} -- \ref{__crit_E}) can be consistently extended into quantum mechanics for systems with internal degrees of freedom in the case of a zero-temperature environment. This finding complements the previous no-thermalization-without-correlations result \cite{2017-Zhdanov} implying that such dissipators are absent for structureless particles. We proved and illustrated on the analytically tractable 
 example that the internal degrees of freedom enable the quantum friction by serving as an ancilla subsystem to harvest the required correlations and mimic system-bath quantum proximity effects. Informally, this implies that in order to be thermodynamically consistent, quantum friction must dissipate heat both into an environment and inside the system itself. For this to be true, external and ancilla degrees of freedom need to be non-adiabatically coupled.

Quantum friction can be used as a simple phenomenological relaxation model to simulate the non-equilibrium dynamics of complex molecular systems strongly coupled to an homogeneous environment (e.g., a molecule in a solvent). Such a model is guaranteed to be consistent in the thermodynamic limit and may allow for substantial memory and time savings in numerical studies of fundamental photoinduced processes, such as photoisomerization, light-induced charge and energy transfer in organic materials.

The existence of friction-like quantum dissipators at finite temperatures remains an intriguing open question. The affirmative or negative answers would challenge the microscopic or semiclassical theories of friction. For computational applications permitting approximate thermalization, the relaxed thermalization workaround \eqref{01.-L_therm_approx} may be used in place of taxing microscopic models \cite{BOOK-Jacobs,BOOK-Alicki}, if system-bath couplings are weak.

\appendix

\section{Proof of the inequality \texorpdfstring{\eqref{04.-no-go_App_1}}{}\label{@APP:1}}
The proof is by contradiction. Suppose that the inequality \eqref{04.-no-go_App_1} can be turned into equality. This assumption implies that $\matel{n}{\Lbd_{\hat A_k}[\rhoth{\theta}]}{l}{=}0$ for all $n$ an $l$ or, more explicitly, that
\begin{gather}\label{02.-R-S}
\forall n,l: R_{n,l}{-}S_{n,l}{=}0,
\end{gather}
where
\begin{align}\notag
R_{n,l}&{=}\sum_m\gamma_m(\theta)\scpr n{\varphi_{k,m}}\scpr{\varphi_{k,m}}l,\\\notag
S_{n,l}&{=}\scpr{\varphi_{k,n}}{\varphi_{k,l}}\tfrac{\gamma_l(\theta){+}\gamma_{n}(\theta)}2.
\end{align} 
Here $\gamma_i(\theta){=}\matel{i}{\rhoth{\theta}}{i}{=}\frac{e^{-E_i/\theta}}{\sum_ne^{-E_n/\theta}}$ is a positive statistical weight of the $i$-th eigenstate.

The equalities \eqref{02.-R-S} imply that $\sum_{n,l}|R_{n,l}|^2{-}\sum_{n,l}|S_{n,l}|^2{=}0$. After some algebra, this relation reduces to
\begin{gather}
\sum_{m,n}|\scpr{\varphi_{k,m}}{\varphi_{k,n}}|^2\bigg[\frac{\gamma_m(\theta){-}\gamma_{n}(\theta)}2\bigg]^2{=}0.
\end{gather}
Since the term in square brackets is positive for all $m{\ne}n$, we can conclude that the necessary condition for thermalization is that 
$\scpr{\varphi_{k,m}}{\varphi_{k,n}}{=}0$, or $\matel{m}{\hat A_k^{\dagger}\hat A_k}{n}{=}0$ for all $m{\ne}n$. The latter implies that 
\begin{gather}\label{02.-A^+A.ket{m}=c_k.ket{m}}
\forall m: {\hat A_k^{\dagger}\hat A_k}\ket{m}{=}{\hat f_k^{\dagger}\hat f_k}\ket{m}{=}c_{k,m}\ket{m},
\end{gather}
where $c_{k,m}$ are certain nonnegative constants. The nonnegative Hermitian operator ${\hat A_k^{\dagger}\hat A_k}$ can be expanded as
\begin{gather}\label{02.-A^+A-def}
{\hat A_k^{\dagger}\hat A_k}{=}\sum_{i,j{=}0}^{N-1}\proj[\es{i}]{\es{j}}F_{i,j}(\hat\pp),
\end{gather}
where $F_{i,j}(p)$ are some complex-valued functions, such that $F_{i,j}(p){=}F_{j,i}^*(p)$. Using Eq.~\eqref{02.-A^+A-def}, the equality \eqref{02.-A^+A.ket{m}=c_k.ket{m}} can be rewritten in the matrix form
\begin{gather}
\sum_jF_{i,j}(\pp)\Psi_{m,j}(\pp){=}c_m\Psi_{m,i}(\pp).
\end{gather}
Note that here we can treat $\pp$ as c-numbers. The $\Nint{\times}\Nint$ matrix $F$ has at most $\Nint$ distinct eigenvalues $\lambda_k$. Each eigenstate $\ket n$ is associated with one of these eigenvalues $\lambda(n)$. Let us choose the set of $\Nint$ indices $r_k$, such that $\ket{r_k}$ is associated with $\lambda_k$. Then, each of remaining eigenstates should be representable as a linear combination of $\ket{r_k}$,
\begin{gather}\label{02.-cstr1}
\Psi_{m,i}(p){=}\sum_{k:\lambda(r_k){=}\lambda(m)}c_{m,r_k}(\pp)\Psi_{r_k,i}(\pp),
\end{gather}
with $\pp$-dependent coefficients $c_{r_k,m}(\pp)$. Furthermore,
\begin{gather}\label{02.-cstr2}
\lambda(n){\ne}\lambda(m) \Rightarrow \sum_i\Psi_{n,i}^*(\pp)\Psi_{m,i}(\pp){=}0 \mbox{ for all }\pp.
\end{gather}
The basis states $\ket n$ can generally satisfy constraints \eqref{02.-cstr1} and \eqref{02.-cstr2} only in two cases: 1) some of the internal states are decoupled from the rest (i.e., the dynamic space splits into isolated subspaces) and 2) all $\lambda_k$ are equal. Here we are not interested in case 1) and assume that the dynamics of all the external and internal degrees of freedom is mixed by couplings. Case 2) implies that
\begin{gather}
{\hat A_k^{\dagger}\hat A_k}{=}\lambda{\geq}0,
\end{gather}
where $\lambda$ is some nonnegative constant. Without loss of generality, it is sufficient to consider two cases: $\lambda{=}0$ and $\lambda{=}1$. 

The case $\lambda{=}0$ implies that $\forall m:\hat A_k\ket{m}{=}0$, which can be satisfied only if $\hat A_k{=}0$. Hence, we can exclude the case $\lambda{=}0$ from consideration.

Consider now $\lambda{=}1$. In this case, the equality $\Lbd_{\hat A_k}[\rhoth{T}]{=}0$ takes the form
\begin{gather}
\hat A_k\rhoth{T}\hat A_k^{\dagger}{-}\rhoth{T}{=}0.
\end{gather}
It is easy to show that in the case of non-degenerate $\gamma_k(\theta)$ the above equality can be satisfied iff $\hat A_k{=}1$, i.e., when $\Lbd_{\hat A_k}{=}0$. This completes the proof.

\acknowledgments
D. V. Zh. thanks Madonna and Sameer Patwardhan for inspirational discussions broadly exceeding the scope of this work. T. S. and D. V. Zh. thank the National Science Foundation (Award number CHEM-1012207 to T. S.) for support. D. I. B. is supported by AFOSR Young Investigator Research Program (FA9550-16-1-0254).

\bibliography{nD_thermalization}